\documentclass[%
 reprint,
 amsmath,amssymb,
 aps,
]{revtex4-2}

\usepackage{graphicx}
\usepackage{dcolumn}
\usepackage{bm}
\usepackage{physics}
\usepackage{appendix}
\usepackage{hyperref}
\hypersetup{ 
    colorlinks=true,
    linkcolor=blue,     
    urlcolor=blue,
    citecolor=blue,
    }
\usepackage{xspace}

\newcommand{\vect}[1]{{\bm{#1}}}
\newcommand{\vectr}{\bm{r}}
\newcommand{\vectrprime}{\bm{r^{\prime}}}
\newcommand{\phistarmr}{\phi_m^*(\vectr)}

\newcommand{\phimmoner}{\phi_{m-1}(\vectr)}
\newcommand{\phimr}{\phi_m(\vectr)}
\newcommand{\phistarm}{\phi_m^*}
\newcommand{\phim}{\phi_m}
\newcommand{\phistar}{\phi^*}

\newcommand{\fnr}{{\left(\vect{r}\right)}}

\newcommand{\vectk}{\bm{k}}

\begin{document}

\preprint{APS/123-QED}

\title{Finite temperature stability of quantized vortex structures in rotating Bose-Einstein condensates via complex Langevin simulation}

\author{Kimberlee Keithley$^1$}
\author{Kris T. Delaney$^2$}%
\author{Glenn H. Fredrickson$^{1,2}$}
\affiliation{$^1$Department of Chemical Engineering, University of California, Santa Barbara, California 93106, USA\\
$^2$Materials Research Laboratory, University of California, Santa Barbara, California 93106, USA
}

\date{\today}

\begin{abstract}
The thermodynamic stability of quantized vortex patterns in rotating Bose-Einstein condensates is assessed at finite temperature using complex Langevin sampling. We construct a temperature-rotation frequency phase diagram and find that that vortices are stabilized at lower rotation speeds by the addition of quantum and thermal fluctuations. The coherent states field theoretic representation of the imaginary time path integral enables efficient simulation of large systems at finite temperature, and the complex Langevin simulation scheme bypasses the sign problems that arise from the complex-valued coherent states fields as well as the gauge potential describing solid body rotation. Field operators allow us to generate high-resolution images of particle and momentum density of the cloud. Quantized vortices appear as dark spots on density images, and vector plots of cloud momentum detail circulation around each vortex.
\end{abstract}

\maketitle

\section{Introduction}
It has been known since the discovery of superfluidity in He-4 below 2.2K~\cite{kapitza, allen_and_misener} that mechanical rotation of a superfluid gives rise to quantized vortices, so-named because their circulation is discretized in units of $h/m$. Small numbers of vortices have been observed in stirred dilute Bose-Einstein condensates~\cite{matthews_vortices_1999, madison_vortex_2000}, and have been induced in unstirred gas clouds during quenching through the transition temperature~\cite{weiler_spontaneous_2008} and in rotating normal gases near the critical point~\cite{haljan_driving_2001}.   
Giant Abrikosov arrays of quantized vortices in dilute harmonically trapped BECs have also been observed~\cite{abo-shaeer_observation_2001}, and there is some experimental work investigating low-lying excitations close to the centrifugal limit~\cite{schweikhard_rapidly_2004}. More recent work has gone beyond the centrifugal limit by imposing a quartic trapping potential~\cite{bretin_fast_2004}.

However, experiments that probe further into high-rotation regimes pose difficulties both in implementation and imaging. Theoretical analyses predict the melting of the giant Abrikosov lattice array at a fill fraction, $\nu = N_\mathrm{particles}/N_\mathrm{vortices}$, of around 10~\cite{cooper_quantum_2001} and at around $\nu=1$, researchers have suggested the emergence of complex quantum Hall states; for example, a Laughlin ground state with no long-range off-diagonal order and vanishing short-range interactions at $\nu = 1/2$~\cite{fetter_rotating_2009, cooper_composite_1999}. However, there are no exact $T=0$ results for $\nu \lesssim 10$~\cite{bloch_many-body_2008}.

Thus far, numerical simulations based on the Gross-Pitaevskii equation have been used to great success in quantitatively and qualitatively understanding both time-dependent \cite{simula_giant_2004, zhang_dynamics_2007, villois_evolution_2016, amette_estrada_turbulence_2022, tsubota_dynamics_2003, tsubota_vortex_2002, sinha_dynamic_2001, saito_split-merge_2004, wang_quantized_2011, barnett_vortex_2010, kasamatsu_three-dimensional_2005, kasamatsu_vortex_2009, kasamatsu_giant_2002, lundh_multiply_2002, saito_emergence_2004, tengstrand_rotating_2019} and equilibrium \cite{aftalion_vortex_2001, aftalion_vortex_2005, aftalion_giant_2004, aftalion_shape_2002, aftalion_three-dimensional_2003, bao_ground_2005, castin_bose-einstein_1999, cooper_quantum_2001, jackson_phase_2004, kasamatsu_vortex_2003, mason_classification_2011, kavoulakis_weakly_2000} properties of quantized vortices in systems of rotating BECs, neglecting quantum and thermal fluctuations.
Extended-GPE methods, which couple the GPE to a semi-classically described non-condensate band, such as stochastic GPE and the Zaremba-Nikuni-Griffin model, also give some insights into the general effect of fluctuations \cite{mink_vortex-lattice_2009, weiler_spontaneous_2008, simula_thermal_2006, jackson_finite-temperature_2009}, but ultimately the description is approximate in nature.
Analytical perturbation theories have provided additional information into the effect of low-lying excitations ~\cite{baym_vortex_2004, duine_noisy_2004, duine_stochastic_2001, fischer_vortex_2003, jackson_vortices_2004, kavoulakis_rapidly_2003, kavoulakis_rotating_2004, kim_dynamics_2005, mizushima_collective_2004, stringari_phase_1999, ho_bose-einstein_2001}.
However, exact treatment of both quantum and thermal fluctuations remains elusive. 

Path integral Monte Carlo, which treats quantum and thermal fluctuations exactly, is hindered by the presence of an irremovable sign problem in the Hamiltonian due to the gauge potential describing rotation about the $z$-axis.
One possible resolution to the sign problem is to use simplifying approximations to remove the problematic degrees of freedom.
For example, in the rotating system the complex phase can be removed by making a ``fixed phase'' approximation~\cite{galambos_path-integral_2018}.
Another possible resolution is to employ a different sampling method that does not place restrictions on the sign of the statistical weight.

In this work, we employ the complex Langevin sampling method, a fictious dynamics scheme for field theoretic formulations of the partition function~\cite{parisi, klauder} that is robust for actions with a sign problem.
Because there is no Monte Carlo importance sampling, there is no restriction on the sign of the statistical weight.
This method has seen success over the years in many areas of physics, such as polymer science~\cite{delaney_recent_2016, delaney_theory_2017, lee_complex_2008, man_coherent_2014}, lattice QCD~\cite{aarts_qcd_2016, aarts_stochastic_2008, ito_complex_2020, joseph_complex_2021, kogut_applying_2019, sexty_simulating_2014}, and more recently, in cold atom systems both Bose~\cite{delaney_numerical_2020, mcgarrigle_emergence_2023, heinen_complex_2022, heinen_simulating_2023, hayata_complex_2015, berger_interacting_2019, attanasio_thermodynamics_2020} and Fermi~\cite{attanasio_pairing_2022, loheac_third-order_2017, rammelmuller_finite-temperature_2018}.
In particular, there has been one recent attempt to simulate quantized vortices in rotating BECs by Hayata and Yamamoto~\cite{hayata_complex_2015} which demonstrated the viability of the method, but did not map in detail vortex transitions, or visualize vortex structures.

In this study, we numerically construct a phase diagram mapping the stability of the rotating BEC with different vortex structures as a function of angular velocity at select non-zero temperatures below the critical transition.
By leveraging a new method of accessing free energies in field theoretic methods~\cite{fredrickson_direct_2022}, we are able to directly calculate the free energy of the system at a given temperature and angular velocity.
We confirm the existence of vortex cores by plotting high-resolution images of cloud density, and confirm circulation of particles around vortex cores by visualizing the momentum vector field.

\section{Numerical Methods}
\subsection{Coherent states field theory}\label{sec:theory}
We begin with the second quantized Hamiltonian for non-relativistic bosons confined in an anisotropic harmonic trap and rotating counterclockwise about the $\hat{\vect{z}}$ axis at a fixed rotational frequency $\Omega$ with a frame of reference rotating with the walls of the trap,
\begin{multline}\label{eqn:hamiltonian}
    \hat{H} = \int d^d r\, \hat{\Psi}^{\dagger}(\vectr)\left(\frac{-\hbar^2}{2 m}\laplacian +U_\mathrm{ext}(\vectr)-\mu \right. \\
    \left.\frac{}{} - \Omega i \hbar (y \partial_x - x \partial_y)\right)\hat{\Psi}(\vectr) + \frac{g}{2} \hat{\Psi}^{\dagger}(\vectr)  \hat{\Psi}^{\dagger}(\vectr) \hat{\Psi}(\vectr) \hat{\Psi}(\vectr).
\end{multline}
Here, $U_\mathrm{ext}(\vectr) = \frac{1}{2}m\omega_{xy}^2(x^2 + y^2 + \gamma z^2)$ where $\gamma = \omega_z^2/\omega_{xy}^2$ and $\omega_{xy}$ and $\omega_z$ are  frequencies of trap confinement in the $xy$ plane and along $z$, respectively.
The chemical potential is denoted $\mu$ and the contact interaction strength is denoted $g = 4 \pi \hbar^2 a_s /m$, with $a_s$ the experimental s-wave scattering length and $m$ the mass of the atom.

By the usual method~\cite{Orland_book}, we then construct the imaginary time path integral representation of the many body problem. Rather than particle coordinate vectors $\vectr^n$, we use complex-conjugate coherent states fields $\phi_m(\vectr)$ and $\phi_m^*(\vectr)$ as the basis, where $m$ is a discrete imaginary time index $m \in \{0, 1, ..., M-1\}$, with M corresponding to the number of imaginary time slices.
The grand canonical partition function in this coherent states basis is
\begin{equation}\label{eqn:GCE_pf}
    \mathcal{Z}_G =  \prod_{m=0}^{M-1} \left[\int_{\phi_{M}(\boldsymbol{r}) = \phi_{0}(\boldsymbol{r})} \mathcal{D}(\phi_m^*, \phi_m) \right] e^{-S \left[ \phi, \phi^* \right]}.
\end{equation}
Note that periodic boundary conditions in imaginary time are enforced by the requirement $\phi_0(\vectr) = \phi_M(\vectr)$.

The form of the action functional $S$ for this system is
\begin{widetext}
\begin{multline}\label{eqn:GCE_action}
S = \epsilon \sum_{m=0}^{M-1} \int d^d r  \: \phistarm\left[\frac{\phi_{m}\left(\vect{r}\right)-\phi_{m-1}\left(\vect{r}\right)}{\epsilon} + \left( -\frac{\hbar^2}{2m} \laplacian +
    U_\mathrm{ext}(\vect{r}) - \mu - \Omega i \hbar (y \partial_x - x \partial_y) \right) \phi_{m-1}(\vect{r})  \right]\\
    + \frac{g \epsilon}{2} \sum_{m=0}^{M-1}\int d^d \vectr \, \left[\phistarm(\vectr)\right]^2 \left[\phi_{m-1}\left(\vect{r}\right)\right]^2,
\end{multline}
\end{widetext}
with $\epsilon = \beta / M$ and $\beta=1/k_B \mathrm T$.

Thermodynamic observables are calculated by ensemble averaging coherent states field operators.
Operators are derived by taking appropriate partial derivatives of the partition function, Eqn.~\ref{eqn:GCE_pf}.
For a given observable $O$, if there exists a corresponding coherent states field operator $\tilde{O}[\phi, \phistar]$, then $O = \langle \tilde{O}[\phi, \phistar] \rangle$.
The angle brackets denote ensemble averages with complex weight $\exp(-S)$.

Some examples of useful field operators~\cite{G_K_book} include the internal energy,
\begin{multline}\label{eqn:internalenergyoperator}
    \tilde{E}[\phi, \phi^*] = \frac{1}{M} \sum_{m=0}^{M-1} \int d^d r \,  \left\{ \phistarmr \left( \frac{-\hbar^2}{2m} \nabla^2 + {U}_\mathrm{ext}(\vectr) \right. \right. \\
    \left. \left. \frac{}{} - \Omega i \hbar (y \partial_x - x \partial_y)\right)\phimmoner
     + \frac{{g }}{2} [\phistarmr \phimmoner]^2 \right\},
\end{multline}
and the scalar density field,
\begin{equation}\label{eqn:densityoperator}
    \tilde{\rho}(\vectr)[\phi, \phistar] = \frac{1}{M} \sum_{m=0}^{M-1} \phistarmr \phimmoner.
\end{equation}
Particle number is then obtained by integrating the density field over space,
\begin{equation}\label{eqn:partnumoperator}
    \tilde{N}[\phi, \phistar] = \frac{1}{M} \sum_{m=0}^{M-1} \int d^d r \: \phistarmr \phimmoner.
\end{equation}

Field operators are not limited to only scalar quantities; there also exists a vector operator for momentum density,
\begin{equation}\label{eqn:momentumoperator}
    \tilde{\vect{p}}(\vectr)[\phi, \phistar] = \frac{1}{M}\sum_{m=0}^{M-1} \phistarm(\vectr) \left[-i\hbar \nabla \right] \phi_{m-1}(\vectr).
\end{equation}

Critically, we are also able to derive an operator for free energy in a manner similar to that of~\cite{fredrickson_direct_2022}.
A key difference, however, is that the presence of a trap means that the system boundary is no longer the rigid walls of the box. To take the necessary pressure derivative, we follow the procedure in~\cite{romero-rochin_EOS_2005}, generalized to anisotropic traps.
The resulting Hemholtz free energy field operator is
\begin{multline}\label{eqn:FreeEnergy}
    \tilde{A}[\phi, \phistar] = - \frac{2}{3} \left< \frac{1}{M} \sum_{m=0}^{M-1} \int d^d r \, \phistarmr \frac{1}{2} \omega^2_\mathrm{GA} \right.
    \left.\frac{}{} \right. \\
    \left. \times \left(\frac{\omega_x^2}{\omega^2_\mathrm{GA}} x^2 + \frac{\omega_y^2}{\omega^2_\mathrm{GA}}y^2 + \frac{\omega_z^2}{\omega^2_\mathrm{GA}}z^2 \right) \phimmoner\right> \\
    + N \left< \tilde{\mu}[\phim, \phistarm] \right>,
\end{multline}
where $\omega_\mathrm{GA} = ({\omega_x} {\omega_y}\omega_z)^{1/3}$ is the geometric mean of the components of the trap frequencies.

To reduce the number of relevant parameters, we rescale lengths by $L_c = \sqrt{\hbar /(m \omega_{xy})}$ and energies by $E_c = \hbar \omega_{xy}$.
Inserting into Eqn. \ref{eqn:GCE_action} and simplifying gives
\begin{multline}\label{eqn:ND_action}
S = \bar\epsilon \sum_{m=0}^{M-1} \int d^d r  \: \phistarm\left[\frac{\phi_{m}\left(\vect{r}\right)-\phi_{m-1}\left(\vect{r}\right)}{\bar\epsilon}\right. \\ 
    \left. + \left\{\laplacian + U_\mathrm{ext}(\vect{r}) - \bar\mu - \bar\Omega i (y \partial_x - x \partial_y) \right\} \phi_{m-1}(\vect{r})  \right]\\
    + \frac{\bar g \bar\epsilon}{2} \sum_{m=0}^{M-1}\int d^d r \, \left[\phistarm(\vectr)\right]^2 \left[\phi_{m-1}\left(\vect{r}\right)\right]^2,
\end{multline}
where the bars distinguish dimensionless from dimensional quantities.
We include bars only on the relevant physical parameters; $\bar{\mu} = \mu/E_c$, non-dimensional interaction strength $\bar{g} = g / (E_c L_c^d)$, non-dimensional rotation speed $\bar{\Omega} = \Omega / \omega_{xy}$, non-dimensional temperature $\bar{\mathrm T} = k_B \mathrm{T} /E_c$, $\bar\epsilon = 1/(M\bar{\mathrm T})$, and neglect them on other quantities for brevity.
From here on, all equations are given in non-dimesional form unless otherwise specified.

Operators are non-dimensionalized by the same procedure.
For example, the non-dimensional vector operator for momentum becomes
\begin{equation}\label{eqn:momentumoperatorND}
    \tilde{\vect{p}}(\vectr)[\phi, \phistar] = \frac{1}{M}\sum_{m=0}^{M-1} \phistarm(\vectr) \left[-i \nabla \right] \phi_{m-1}(\vectr).
\end{equation}
where $\phistarmr$ and $\phimmoner$ are non-dimensionalized by scaling by a factor of $L_c^{-d/2}$. 

\subsection{Complex Langevin method}
Because the action in Eqn.~\ref{eqn:GCE_action} is complex-valued, the sign problem prevents efficient sampling with conventional Monte Carlo (MC) techniques.
Instead, we employ a sampling method known as complex Langevin (CL).
In the CL formalism the complex fields $\phi$ and $\phistar$ are allowed to independently evolve stochastically in fictitious CL time.
Based on previous experience with polymer systems~\cite{man_coherent_2014} and success with a homogeneous Bose gas~\cite{delaney_numerical_2020}, we adopt an off-diagonal descent scheme that decouples $\phi$ and $\phistar$ to first order and allows for efficient sampling and stable time integration
\begin{equation}\label{eqn:CL_EOMs_phi}
    \frac{\partial \phim(\vectr, t)}{\partial t} = - \frac{\delta S}{\delta \phistarm(\vectr, t)} + \gamma_m(\vectr, t)
\end{equation}
\begin{equation}\label{eqn:CL_EOMs_phistar}
    \frac{\partial \phistarm(\vectr, t)}{\partial t} = - \frac{\delta S}{\delta \phim(\vectr, t)} + \gamma_m^*(\vectr, t)
\end{equation}
where $t$ is fictitious CL time, and the correlation statistics of $\gamma$ are chosen to satisfy the fluctuation-dissipation theorem~\cite{McQuarrie_book, van_Kampen_book}.
In the present case, $\gamma$ is of the form $\gamma_m(\vectr, t) = \eta_m^{(1)}(\vectr, t)+i\eta_m^{(2)}(\vectr, t)$ where $\eta^{(i)}$ are real-valued Gaussian random variables with covariance $\langle \eta_m^{(i)}(\vectr, t) \eta_n^{(j)}(\vectr^{\prime}, t^{\prime})\rangle = \delta_{ij} \delta_{mn}\delta(\vectr - \vectrprime)\delta(t-t^{\prime})$ and $\gamma_m^*(\vectr, t)$ is the complex conjugate of $\gamma_m(\vectr, t)$.

Additionally, because experimental realizations of such systems are typically performed at fixed particle number rather than fixed chemical potential, we reformulate Eqn.~\ref{eqn:GCE_action} into an expression for the canonical partition function.
We first insert a constraint into Eqn.~\ref{eqn:GCE_pf}~\cite{delaney_numerical_2020},
\begin{equation}
        \mathcal{Z}_c = \int \mathcal{D}(\phistar, \phi) \: \delta(N-\tilde{N}[\phistar, \phi]) e^{-S_0}
\end{equation}
where $S_0$ is the action in Eqn.~\ref{eqn:GCE_action} with $\bar \mu = 0$.
By inserting an exponential representation of the delta function we obtain an effective canonical partition function
\begin{equation}\label{eqn:CE_PF}
    \mathcal{Z}_c = \frac{1}{2 \pi} \int dw \int \mathcal{D}(\phistar, \phi) e^{-iw(N-\tilde{N}[\phi, \phistar])} e^{-S_0},
\end{equation}
where $\tilde{N}$ is the coherent-states field operator for particle number, Eqn.~\ref{eqn:partnumoperator}, $N$ is the nominal particle number of the canonical ensemble, and $w$ is a real-valued scalar.
The physical meaning of $w$ soon becomes apparent by recognizing that the thermodynamic derivative $\partial \ln \mathcal{Z}_c /\partial N$ can be written alternatively as $\beta \mu$ or $\langle iw \rangle$, where the latter is an average taken with the statistical weight of Eqn.~\ref{eqn:CE_PF}.
This allows us to write a Langevin equation of motion for a time-dependent complex-valued chemical potential, $\bar\mu(t)$, that upon averaging yields the real-valued chemical potential of a system of $N$ particles:
\begin{equation}\label{eqn:constraint}
    \partial_t {\bar\mu}(t) = \bar{\mathrm T} \left[  \lambda_{{\mu}}(N - \tilde{N}[\phi, \phistar]) +  i\eta_{{\mu}}(t) \right],
\end{equation}
where $\eta_{\mu} (t)$ is real-valued Gaussian noise with $\langle \eta_{\mu}(t) \eta_{\mu}(t^{\prime})\rangle = 2 \lambda_{\mu} \delta(t - t^{\prime})$ and $\lambda_\mu$ is a relaxation coefficient that can be adjusted to control the rate of chemical potential updates relative to field updates.

It can be proven that if the stochastic equations of motion in Eqns.~\ref{eqn:CL_EOMs_phi},~\ref{eqn:CL_EOMs_phistar} and~\ref{eqn:constraint} reach a time-independent solution then averages of field operators over CL time are formally equivalent to thermodynamic ensemble averages~\cite{lee_convergence_nodate, gausterer_mechanism_1993}.
In practice, this amounts to simulating the CL equations of motion until a time-independent solution is reached, after which the equations are used to generate a Markov chain of field configurations.
Field operators of interest are evaluated using these field configurations.
The operator ``samples'' from independent runs after CL equilibration are then averaged together, giving an estimate for the observables of interest.
Although instantaneous values of a coherent states field operator $\tilde{O}[\phi, \phistar]$ may be complex-valued, CL averages yield real-valued results for all physical quantities.
Because $e^{-S}$ is never directly calculated as a probability weight, the complex-valued nature of $S$ does not pose a sign problem as it does in Monte Carlo importance sampling.

\subsection{Forward time propagation}
To efficiently simulate Eqns.~\ref{eqn:CL_EOMs_phi} and~\ref{eqn:CL_EOMs_phistar} we employ the pseudospectral method detailed in Delaney et al~\cite{delaney_numerical_2020}. We begin by Fourier transforming the two equations of motion using the Fourier convention $g_{\vectk} = V^{-1} \int d^d r \: f(\vectr) \exp(-i \vectk \cdot \vectr)$ with $k = 2 \pi L^{-1}(l, m, n)$, $l, m, n \in \mathbb{Z}$ for a cubic simulation cell of length $L$.
We then transform to Matsubara frequency by the convention $g_j = M^{-1} \sum_{m=0}^{M-1} f_m \exp(-i m \omega_j /(\bar{\mathrm T} M))$ where $\omega_j = 2 \pi j \bar{\mathrm T}$, $j \in \mathbb{Z}$. The resulting set of equations is
\begin{multline}
\small
    \partial_t \Psi_{j,\vect{k}} = -A_{j,\vect{k}}\Psi_{j,\vect{k}} + \mathcal{F}_2\left[\gamma_m\left(\vect{r}\right)-\bar\epsilon \left\{ w_m\left(\vect{r}\right)\right. \right. \\
    + \left. \left. \bar\Omega i (y \partial_x- x\partial_y) \right\} \phi_{m-1}\left(\vect{r}\right)\right]
\end{multline}
\begin{multline}
\small
    \partial_t \Psi^*_{j,\vect{k}} = -A^*_{j,\vect{k}}\Psi^*_{j,\vect{k}} + \mathcal{F}_2\left[\gamma^*_m\left(\vect{r}\right)-\bar \epsilon \left\{ w_{m+1} \left(\vect{r}\right)\right. \right. \\
    - \left. \left.\bar\Omega i (y \partial_x- x\partial_y) \right\} \phi^*_{m+1}\left(\vect{r}\right)\right]
\end{multline}
with $w_m\left(\vect{r}\right) = \bar{g} \phi_{m-1}\left(\vect{r}\right)\phi^*_m\left(\vect{r}\right) + U_\mathrm{ext}\left(\vect{r}\right)$, $A_{j,\vect{k}} = 1-\left(1-\bar\epsilon k^2/2 + \bar\epsilon \bar{\mu}\right)e^{-2\pi i j/M}$, and $A^*_{j,\vect{k}} = 1-\left(1-\bar\epsilon k^2/2 + \bar\epsilon\bar{\mu}\right)e^{2\pi i j/M}$.
$\mathcal{F}_2$ denotes the double transform to reciprocal space and Matsubara frequency, and the shorthand $\mathcal{F}_2(\phimr) = \Psi_{j,\vect{k}}$ and $\mathcal{F}_2(\phistarmr) = \Psi^*_{j,\vect{k}}$ is used.
Note that $\Psi_{j,\vect{k}}$ and $\Psi^*_{j,\vect{k}}$ are independently fluctuating complex fields, and $A_{j,\vect{k}}$ and $A^*_{j,\vect{k}}$ are not necessarily complex conjugates due to the complex-valued, time-dependent $\bar{\mu}$.

We propagate the system forward in fictitious time by a discrete timestep $\Delta t$ using an exponential time differencing method (ETD1).
We take the linear term with coefficient $A_{j,\vect{k}}$ as an integrating factor over the integration from $t$ to $t+\Delta t$, and the remaining terms are taken at the previous time, $t$, resulting in an algorithm with weak first-order accuracy and an excellent balance of accuracy and stability among first-order integration schemes~\cite{villet_efficient_2014,delaney_numerical_2020}.
It is worth noting that in the absence of a nonlinear term (i.e. $w_m = 0$), forward integration by ETD1 is analytically exact.

The resulting set of equations for the double-Fourier transformed coherent states fields at the advanced timestep are
\begin{multline} \label{eqn:CS-CL_timestepped1}
    \Psi_{j,\vect{k}}^{(t+\Delta t)} \approx e^{-A_{j,\vect{k}}\Delta t}\Psi_{j,\vect{k}}^{(t)} - \left[\frac{1-e^{-A_{j,\vect{k}}\Delta t}}{A_{j,\vect{k}}}\right] \\
   \times \mathcal{F}_2\left[\epsilon w_m^{(t)}\fnr \phi_{m-1}^{(t)}\fnr\right] + R_{j,\vect{k}}^{(t)} \\
\end{multline}
\begin{multline}\label{eqn:CS-CL_timestepped2}
    \left(\Psi^*_{j,\vect{k}}\right)^{(t+\Delta t)} \approx e^{-A^*_{j,\vect{k}}\Delta t} \left(\Psi^*_{j,\vect{k}}\right)^{(t)} - \left[\frac{1-e^{-A^*_{j,\vect{k}}\Delta t}}{A^*_{j,\vect{k}}}\right] \\
    \times \mathcal{F}_2\left[\epsilon w_{m+1}^{(t)}\fnr \left(\phi^{*}_{m+1}\right)^{(t)}\fnr\right] + \left(R^*_{j,\vect{k}}\right)^{(t)}.
\end{multline}
where the superscript in parenthesis indicates whether the quantity indicated is taken at the current time $t$ or the advanced timestep $t+\Delta t$, and $R$ denotes the integrated noise variables.
The proper noise correlations are calculated by integrating the original noise correlations of $\eta$ and $\eta^*$ over the discrete time interval, giving 
\begin{multline}
    \left<R_{j,\vect{k}}^{(t)}R_{j^\prime,\vect{k}^\prime}^{(t^\prime)}\right> =   V^{-1}M^{-1}\delta_{\vect{k},-\vect{k}^\prime}\delta_{t,t^\prime}\delta_{j,-j^\prime} \\
    \times \left[1-\exp\left(-2A_{j,\vect{k}}\Delta t\right)\right] /(2 \Delta t A_{j,\vect{k}})
\end{multline}

We step the constraint, Eqn. \ref{eqn:constraint}, by a simple first order Euler-Maruyama time step:
\begin{equation} \label{eqn:mu_bar}
    \bar\mu^{(t + \Delta t)} =  \bar{T} \lambda_{\mu} (N - \tilde{N}[\phi, \phistar]) \Delta t + \bar\mu^{(t)}
\end{equation}
where we have excluded the purely imaginary Langevin noise to enhance numerical stability.
We justify this assumption in more detail in Section \ref{sec:expts}.

Fast Fourier transforms (FFTs) are used to perform the transforms over spatial and imaginary time coordinates, which requires us to specify periodic boundary conditions in space as well as the imaginary time index throughout this study.
The results are not affected by choice of spatial boundary conditions due to the confining potential that leads to an absence of boson density near the edges of the simulation domain.

\section{Noise-free simulation and Gross-Pitaevskii reference}
We begin by solving the set of equations in Eqns.~\ref{eqn:CL_EOMs_phi} and~\ref{eqn:CL_EOMs_phistar} without the noise terms $\gamma, \; \gamma^*$.
We further restrict ourselves to physically-relevant solutions, which are invariant in the imaginary time index $m$ in the absence of noise.
With these simplifying assumptions, the two CL equations of motion are exactly complex conjugates, and can be simplified into one equation -- the so-called Gross-Pitaevskii equation (GPE).
We also rescale lengths and energies as in Sec.\ref{sec:theory}
The final rescaled GPE equation is
\begin{eqnarray}\label{eqn:GPE}
  \left\{  -\frac{1}{2} {\laplacian} + {U}_\mathrm{ext}({\vectr}) - \bar{\mu} - \bar{\Omega} i ( {y} \partial_{x} -  x \partial_{y}) \right\} \phi (\vectr) \nonumber \\
  + \bar{g} \: |{\phi}({\vectr})|^2  \phi (\vectr) = 0
\end{eqnarray}
Moreover, it is apparent that Eqns.~\ref{eqn:CS-CL_timestepped1} and ~\ref{eqn:mu_bar} with the noise terms $R_{j,\vect{k}}$ set to zero constitute a numerical relaxation scheme for developing solutions to the GPE equation.
This simplification enables comparison to the wealth of literature investigating rotating Bose-Einstein condensates using the GPE.

\subsection{2d Gross-Pitaevskii equation results}
To validate our numerical methods for solving the GPE equation, we reproduced two-dimensional (2d) results from a study by Aftalion and Du in 2001~\cite{aftalion_vortex_2001}, which reports the relative stability of various vortex structures with between one and seven vortices.
We set $\bar{g} = 0.413$ and $N=3030$ such that the non-dimensional group $\epsilon = \left( {\hbar^2}/({2 N g m}) \right)^{1/2} = \left( 1/({2 N \bar{g}}) \right)^{1/2}= 0.02$ as in ref.~\cite{aftalion_vortex_2001}.
We also calculate total internal energy for the same five vortex structures as the reference using the operator in Eqn.~\ref{eqn:internalenergyoperator} with $M = 1$. Cloud density profiles are visualized with a modified form of Eqn.~\ref{eqn:densityoperator} with $M=1$.

\begin{figure}
    \includegraphics[width=\columnwidth]{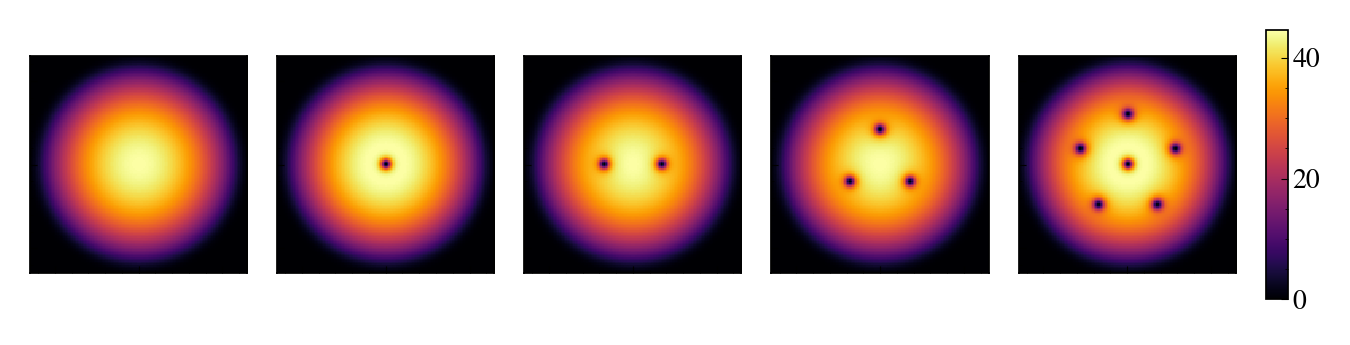}
    \caption{Number density plots show quantized vortices in 2d GPE simulations. From left to right, $\bar{\Omega} = 0.120$, $0.200$, $0.250$, $0.300$, $0.400$. All simulation cells are $24 \cross 24$ non-dimensional length units. The images are cropped to show only the atom cloud.}
    \label{fig:MF_vortices}
\end{figure}

\begin{figure}
    \includegraphics[width=0.45\textwidth]{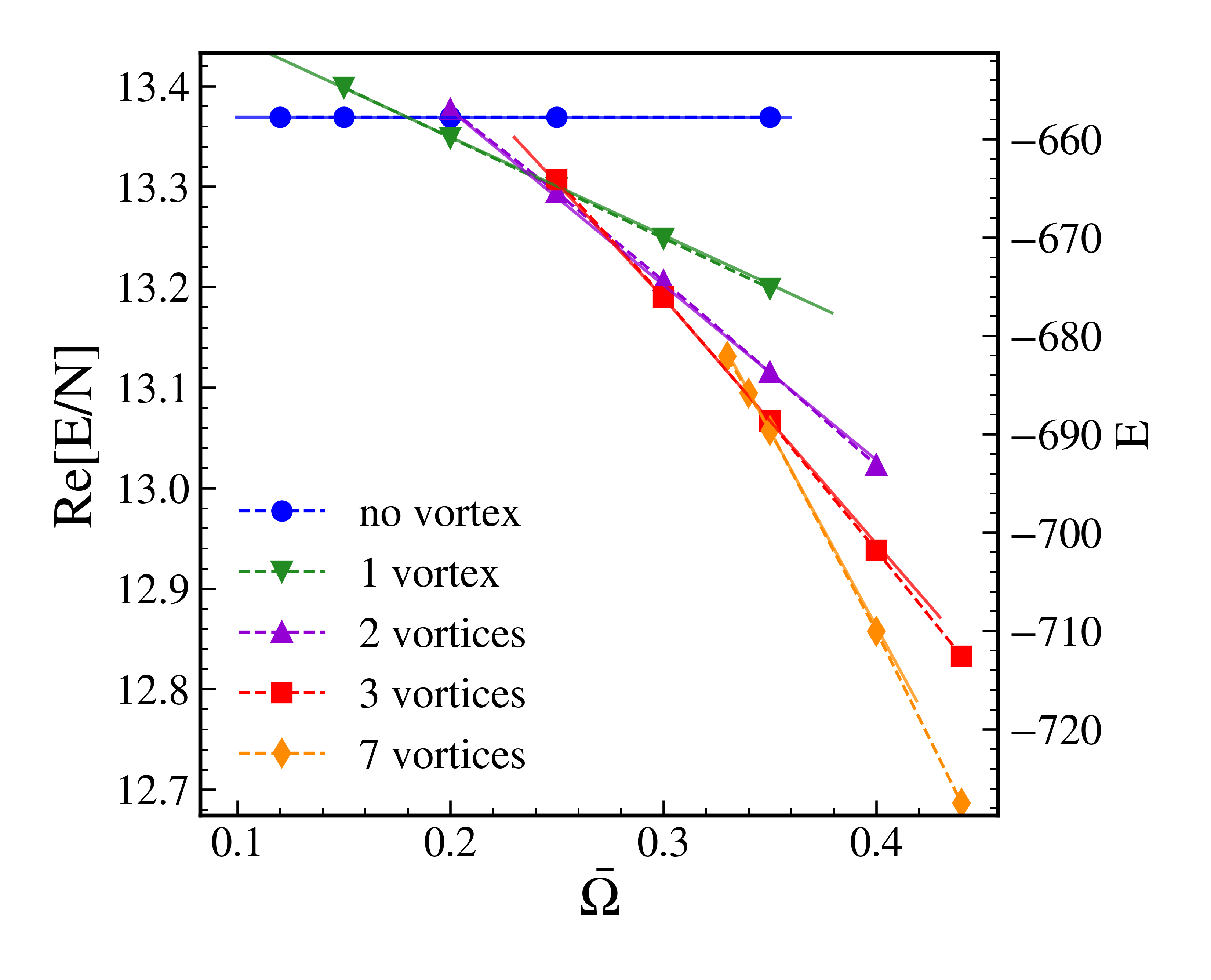}
    \caption{Total internal energy for the vortex configurations shown in Fig.~\ref{fig:MF_vortices}. On the left axis scale, results from complex Langevin simulations (without noise) shown in symbols connected with dashed lines. The dashed lines are a linear fit to the CL data. On the right axis, results for the energy minimizer from reference~\cite{aftalion_vortex_2001} are plotted with solid lines.}
    \label{fig:AD_comparison}
\end{figure}

We first confirm the appearance of quantized vortices visually. Fig~\ref{fig:MF_vortices} shows density plots of five different configurations found in 2d.
Configurations with other numbers of vortices, such as four and five, were also observed, but the energies of these configurations were very close to the energies of the five configurations shown in Fig.~\ref{fig:MF_vortices} and so were omitted for clarity of figures.
We also plot total internal energy (Eqn.~\ref{eqn:internalenergyoperator}) per particle as a function of rotation speed for each configuration (symbols with dashed lines), as well as the energy minimizer reported in Aftalion and Du (solid lines), in Fig.~\ref{fig:AD_comparison}.
We observe excellent agreement in intensive energy between the reference and our simulations over the entire range of rotation speeds, for all five vortex configurations.

\begin{figure}
\includegraphics[width=\columnwidth]{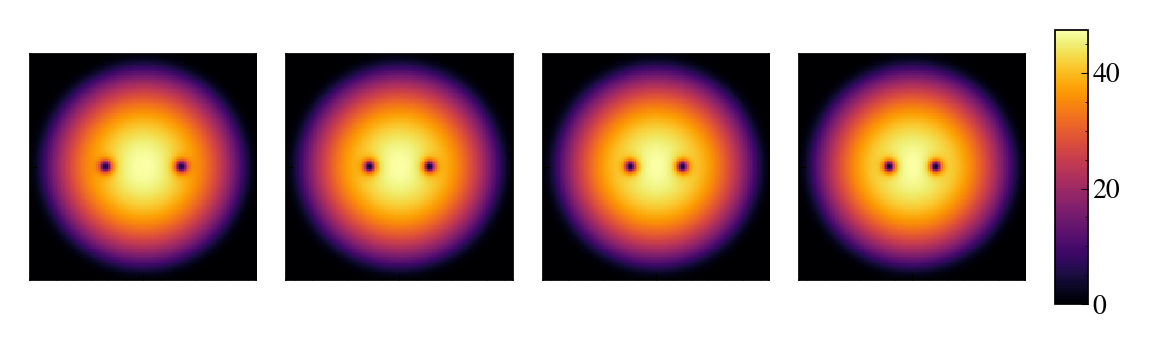}
\caption{Number density plots of 2d GPE simulations with two vortices at different rotation speeds. From left to right, $\bar{\Omega} = 0.200$, $0.250$, $0.300$, $0.350$. Each simulation cell is $12 \cross 12$ non-dimensional length units with $N=3030$ and $\bar{g} = 0.413$. As the rotation speed increases, vortex core spacing decreases.}
\label{fig:vortex_position_2d}
\end{figure}

Within the stability window for a given vortex configuration, the position of vortices is not constant as a function of rotation frequency even with a radially symmetric cloud, as suggested in ref.~\cite{aftalion_vortex_2001}. The $\bar\Omega$ dependence of vortex positions is demonstrated for a two vortex configuration in Fig.~\ref{fig:vortex_position_2d}.
Reference~\cite{castin_bose-einstein_1999} contains an explicit prediction for the relative position of two vortices in a cloud as a function of $\bar \Omega$.
Calculating the vortex self-energy and interaction potential therein for the parameters used in Fig.~\ref{fig:vortex_position_2d} confirms that a decrease in intervortex distance is indeed predicted by the analytical formula.

\subsection{3d Gross-Pitaevskii equation results}
We next proceed to restore the $z$-axis for full 3d simulation. We set $\gamma = 5$ and keep $\bar{g} = 0.413$ for convenience, but increase the particle number to $N = 6,000$ to maintain a cloud width comparable to that of the 2d cloud. We validate the results by verifying the relationship
\begin{equation}\label{eqn:GPE_conservation}
E_\mathrm{\mathrm KE} + E_\mathrm{\mathrm HO} + E_{\Omega} + 2E_\mathrm{\mathrm{INT}}  = N \bar{\mu}
\end{equation}
obtained by direct integration of the Gross-Pitaevskii equation as described in~\cite{P_S_book}, modified to include the contribution of rotation to internal energy, $E_{\Omega} = \int_V d^d r \: \phistar(\vectr) (-\bar{\Omega} i (y \partial_{ x} - {x} \partial_{ y})) \phi(\vectr)$. The remaining contributions on the left-hand side of Eqn.~\ref{eqn:GPE_conservation} from left to right reflect kinetic energy, energy associated with the trap potential, and interaction energy.

\begin{figure}
\includegraphics[width=\columnwidth]{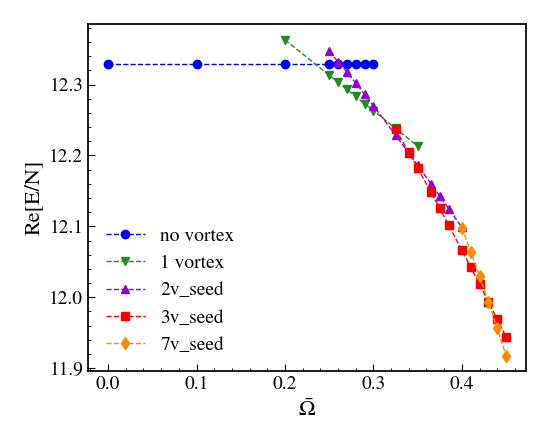}
\caption{Intensive internal energy per particle for the mean-field system with $N=6,000$ and $\bar{g} = 0.413$. Tolerance for convergence of the internal energy was $10 \times 10^{-10}$. Lines are a guide for the eye.}
\label{fig:MF_diagram.png}
\end{figure}

The mean-field phase diagram of the 3d system is presented in Fig.~\ref{fig:MF_diagram.png}. The quantitative structure of the diagram remains much the same despite the fact that holding $\bar{g}$ fixed while increasing dimensionality lowers $g$.
Again, the window of stability for four-, five-, and six-vortex structures is either very narrow, or entirely eclipsed, and we therefore restrict the study to the solutions with zero, one, two, three, and seven vortices.
The internal energy as a function of $\bar{\Omega}$ is nearly linear for each structure, and the globally stable configuration increases from zero to seven sequentially with increasing rotation speed.

\begin{figure}
\raggedright{(a)} \\
    \includegraphics[width=0.24\columnwidth]{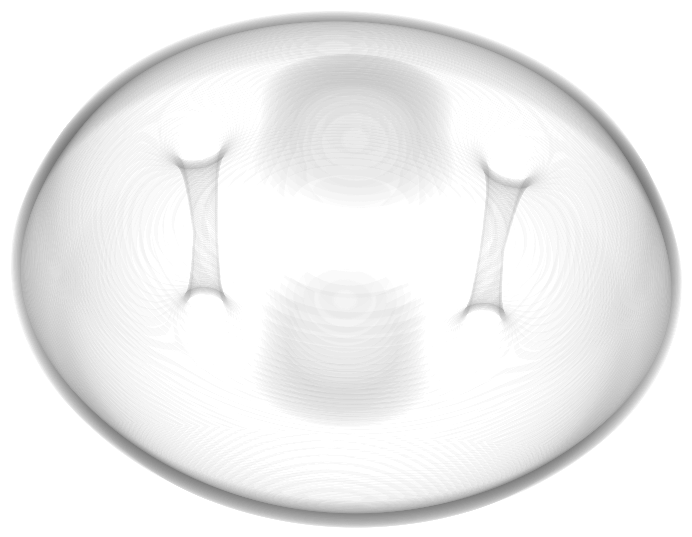}
    \includegraphics[width=0.24\columnwidth]{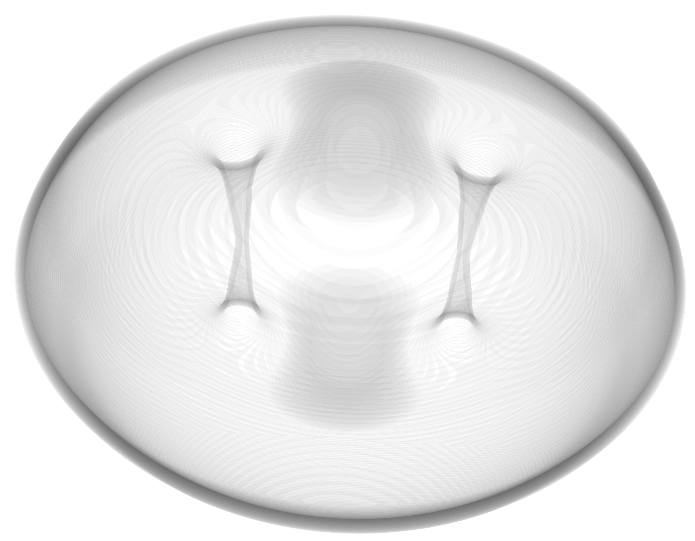}
    \includegraphics[width=0.24\columnwidth]{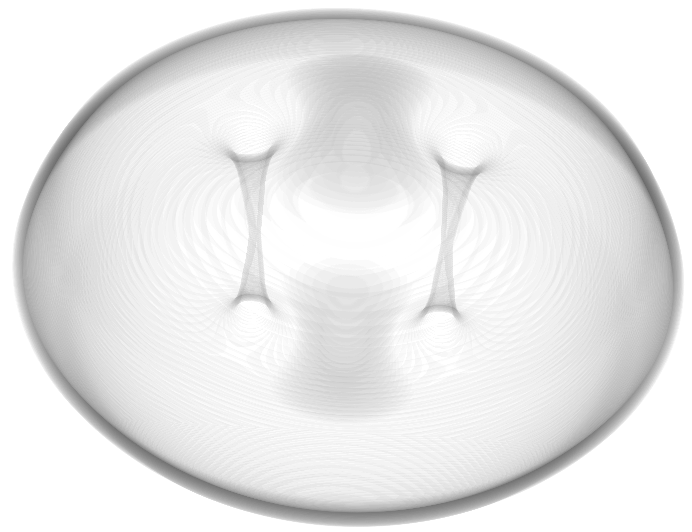}
    \includegraphics[width=0.245\columnwidth]{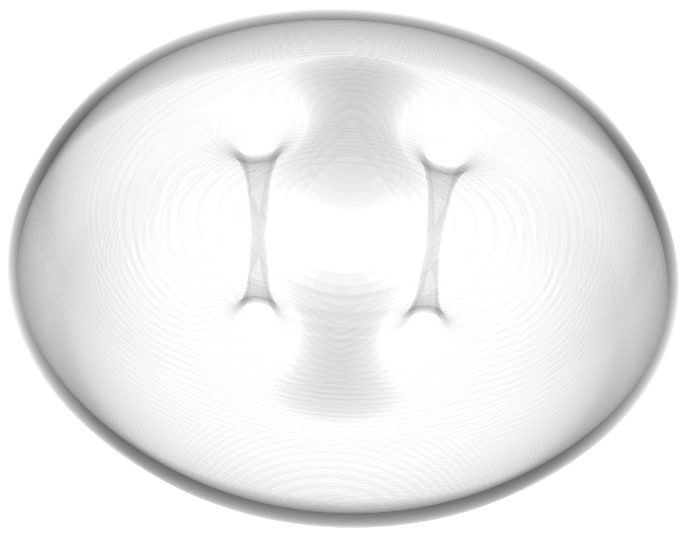}
\\
(b) \hfill
\\
    \includegraphics[width=\columnwidth]{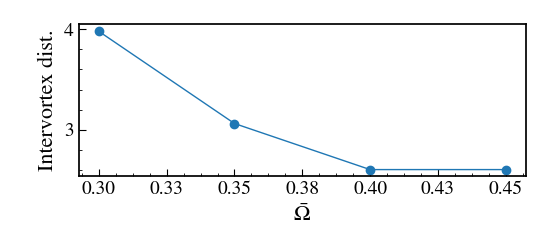}
\caption{(a) Contour plots showing vortex cores in 3d GPE studies. As in the 2d case, increasing rotation speed correlates with decreased vortex spacing. $N=6,000$, $\gamma = 5$, and $\bar{g} = 0.413$, in a simulation cell of 21$\cross$21$\cross$8. From left to right, $\bar{\Omega} = 0.300$, $0.350$, $0.400$, $0.450$.
(b) Intervortex distance as a function of non-dimensional rotation frequency for the 3d mean-field system. Increasing rotation frequency decreases intervortex spacing until about $\bar{\Omega} = 0.400$, after which the spacing stabilizes.}
\label{fig:vortex_position_3d}
\end{figure}

We also find that intervortex distance decreases with increasing $\bar \Omega$.
Contour plots in Fig.~\ref{fig:vortex_position_3d}(a) reveals that vortex lines grow closer as the cloud is rotated faster.
In Fig.~\ref{fig:vortex_position_3d}(b), we measure the intervortex distance and find that the distance stabilizes at above $\bar \Omega = 0.40$, suggesting there is a finite velocity above which the intervortex distance is insensitive to rotation speed.

\section{Complex Langevin Simulation of rotating Bose-Einstein Condensates}\label{sec:expts}
\subsection{Finite temperature phase diagram}
We next restore the noise terms in Eqns.~\ref{eqn:CS-CL_timestepped1}--\ref{eqn:CS-CL_timestepped2}, allowing us to simulate the system at finite temperature with no approximations made to the underlying theory.
The addition of quantum and thermal fluctuations allows the vortex cores to fluctuate around the mean-field solution.
We keep $\bar{g} = 0.413$, $N=6,000$, and $\gamma = 5$.
With the restoration of the noise terms, we also re-introduce temperature, $\bar{\mathrm T}$.
In all studies here, $\bar{\mathrm{T}} / \bar{\mathrm{T}}_c < 1 $ where $\mathrm{\bar T}_c  =  \mathrm{\bar T}_c^0  \left(1 - \bar{\Omega}^2 \right)^{1/3}$ is the transition temperature of the rotating system of non-interacting particles~\cite{stringari_phase_1999}.
The critical temperature of the non-rotating, non-interacting system in non-dimensional units is obtained by the analytical relationship $\bar{\mathrm{T}}_c^0 \approx 0.94 N^{1/3} \gamma^{1/5}$~\cite{bradley_bose-einstein_2008}.
For this section, we fix $\bar{\mathrm{T}} = 5 \approx 0.21 \: \bar{\mathrm{T}}_c^0$.

We find that particle number is converged in preliminary testing in the grand canonical ensemble with $M=48$ imaginary time slices and $\Delta t=0.000250$ CL time resolution and therefore use the same numerical parameters here.
We set the resolution of the $xy$ grid $\Delta_{xy} = 0.123$ with a cell length in the $xy$ plane of $L_{xy} = 14.5$ to accomodate the largest cloud size.
The grid in $z$ remains fixed with a resolution of $\Delta_{z}=1.14$ and cell length $L_{z}=8.00$.

\begin{figure}
    \centering
    \includegraphics[width=\columnwidth]{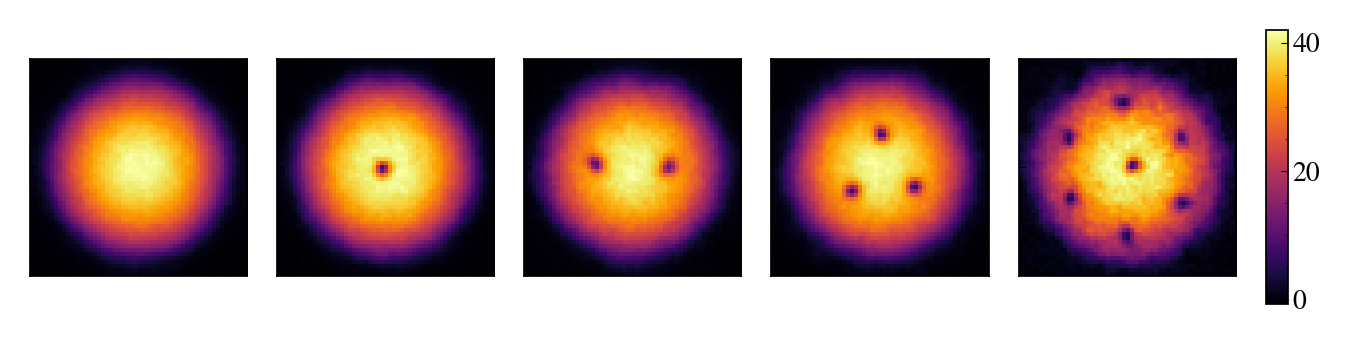}
    \caption{Number density plots of fully fluctuating simulation as seen via a slice through the $xy$ plane. Vortex cores are clearly visible, even with fluctuations. All simulations are performed at $\tilde{\mathrm{T}} = 5$. From left to right, $\bar{\Omega} = 0.000$, $0.100$, $0.200$, $0.350$, $0.400$. The first four images from the left are of simulations performed in a cell of size $12 \cross 12 \cross 8$ non-dimensional length units, and the last simulation at $\bar{\Omega} = 0.450$ is performed in a cell of size $14.5 \cross 14.5 \cross 8$ to account for the expansion of the cloud due to rotation. Images are generated by averaging between 150,000 and 300,000 complex Langevin samples.}
    \label{fig:3d_CL_vortices}
\end{figure}

\begin{figure}
    \includegraphics[width=0.95\columnwidth]{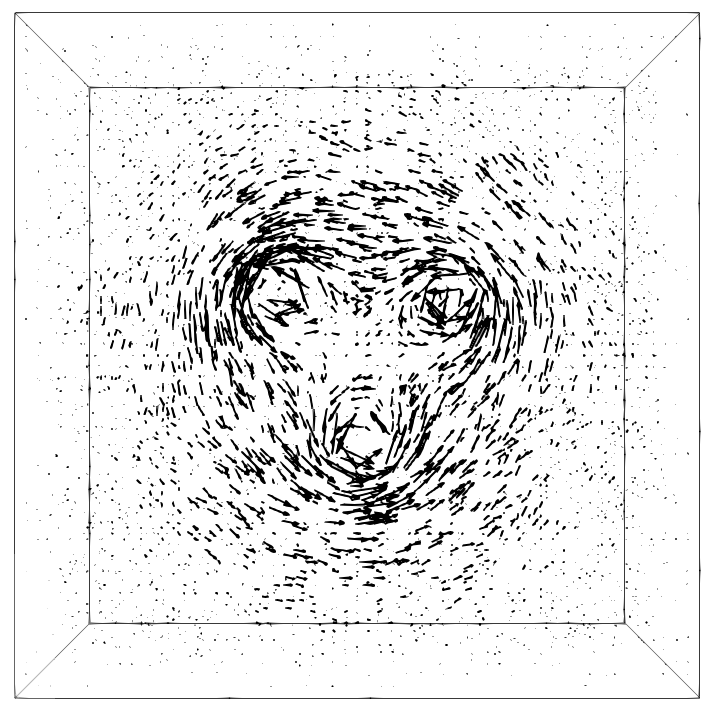}
    \caption{An example of the momentum field operator, Eqn.~\ref{eqn:momentumoperator}, in a cell with three vortices. The system is rotating about the $z$-axis, coming out of the plane of the page, in the counterclockwise direction. The length of the vector arrows is scaled proportional to the magnitude of the momentum at each position. The simulation is run in a cell of size $12 \cross 12 \cross 8$ at $\bar{\mathrm{T}}=5$ and $\bar{\Omega} = 0.350$.}
    \label{fig:momentum_field}
\end{figure}

Using field operators, we obtain high-resolution visual confirmation of quantized vortices.
We first plot the density operator through the $xy$ plane in Fig.~\ref{fig:3d_CL_vortices}.
Because instantaneous snapshots are noisy, we average the density over 150,000--300,000 CL time samples.
All operators are averaged over 100 individual field configurations per CL sample.
Vortices appear as dark spots (density depressions) in the cloud.
The appearance of multi-vortex structures similar to those seen in 2d and 3d mean-field simulation confirms the existence of quantized vortices.
Additionally, the momentum field operator, plotted in Fig.~\ref{fig:momentum_field} for a system with three vortices, confirms circulation around each vortex core.

\begin{figure}
\includegraphics[width=\columnwidth]{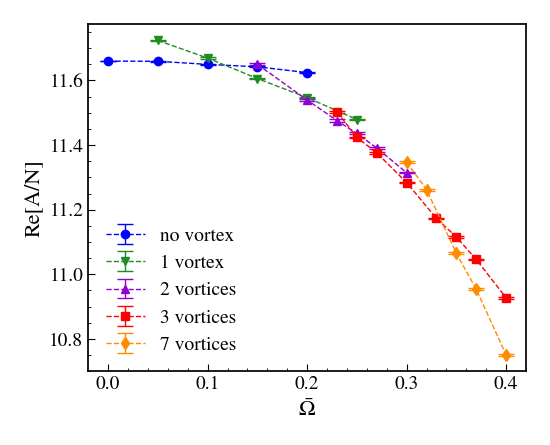}
\caption{Intensive free energy per particle for the fully-fluctuating CL simulations with $\bar{\mathrm{T}} = 5$, $N=6,000$ and $\bar{g} = 0.413$. Free energy is calculated using the operator in Eqn.~\ref{eqn:FreeEnergy}. Some intermediate points are omitted for clarity of the figure. Lines are a guide for the eye.}
\label{fig:NDT5_example}
\end{figure}

Next, we calculate the Helmholtz free energy, Eqn.~\ref{eqn:FreeEnergy}, as a function of $\bar{\Omega}$.
The resulting free energy diagram is shown in Fig.~\ref{fig:NDT5_example}.
From comparison of Fig.~\ref{fig:NDT5_example} and Fig.~\ref{fig:MF_diagram.png}, it is apparent that the addition of fluctuations stabilizes every vortex structure to lower rotation frequencies.

To verify that the exclusion of purely imaginary Langevin noise in the constraint equation, Eqn~\ref{eqn:constraint}, has no effect on the results we run simulations with full noise at $\bar \Omega=0.000$ and $\bar \Omega = 0.400$.
The low-rotation case is seeded with a no-vortex solution obtained by GPE simulation and the high-rotation case is seeded with a 7-vortex solution.
We then compare the average of the internal and Helmholtz free energy operators obtained from the fully-fluctuating simulations to those obtained without noise on the constraint.
In all cases, the difference between operator averages was 0.25\% or less.

Finally, by calculating the difference in free energy between two configurations of interest, $\Delta A = \left<A_1[\phi, \phistar]\right> - \left<A_2[\phi, \phistar]\right>$, and fitting a functional form to the resulting data, we can locate the transition between configurations, denoted $\bar \Omega_c$, at $\Delta A(\bar \Omega_c) = 0$.
The mean-field data are highly linear, but we choose a quadratic equation for fitting the free-energy difference data due to the slight curvature seen in Figure~\ref{fig:NDT5_example}.
Although the choice of quadratic fit is somewhat arbitrary, there was no significant difference between transitions located using linear and quadratic fits.

The calculated transitions are summarized in Table~\ref{tab:transitions}.
Every transition occurs at lower $\bar \Omega$ than the same transition obtained from mean-field calculations, suggesting that fluctuations tend to stabilize vortices.
We further investigate this stabilizing effect in the next section.

\begin{table}
\caption{\label{tab:transitions}%
Transition $\bar{\Omega}$ between vortex structures located by fitting a quadratic equation to the difference of free energies and interpolating.}
\begin{ruledtabular}
\begin{tabular}{ccccc}
$\bar{\mathrm{T}}$ & 0--1 & 1--2 & 2--3 & 3--7 \\ 
\hline
    Mean field & 0.24 & 0.30 & 0.36 & 0.43\\
    5	& 0.12	& 0.18	& 0.27	& 0.31\\
\end{tabular}
\end{ruledtabular}
\end{table}

\subsection{Reduction of critical rotation frequency}
In this section, we focus on the apparent reduction of the critical velocity at which a single vortex appears, $\bar{\Omega}_c$, when fluctuations are added.
To better understand this phenomenon, we first reduce $\bar{g} = 0.0413$, corresponding to a scattering length of about 200$a_0$ for a mass of $^7$Li and $\omega_{xy} \approx 100 \cross 2\pi $ Hz.
We keep $\gamma=5$ and map the transition between the cloud with a vortex and without a vortex at a variety of temperatures.
We set a grid size in the $xy$ direction of $\Delta_{xy} = 0.22$, below the calculated healing length of $\xi \approx 0.26$.
As the results in the previous section did not indicate any vortex bending or other exotic behavior of the vortex lines through the cloud, we resolve only coarsely in the transverse direction to minimize computational cost.
We choose $\Delta_z = 0.8$, the resolution required to obtain a grid-size-independent estimate of internal energy and satisfy the criteria in Eqn.~\ref{eqn:GPE_conservation}.
We keep $M = 48$ as in the previous section.

\begin{figure}
    \includegraphics[width=\columnwidth]{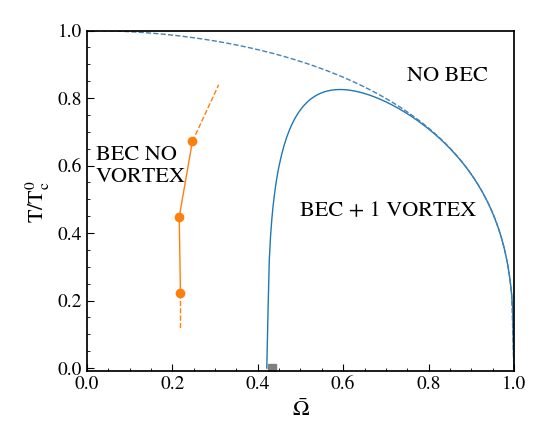}
    \caption{Phase diagram for the anisotropic trapped interacting gas. Solid blue line: analytically predicted transition between the BEC with vortices and BEC without vortices ref~\cite{stringari_phase_1999}. Orange filled circles: Simulation results for $\bar{\Omega}^*$ as a function of temperature. Lines are a guide for the eye. Dashed blue line: Analytically predicted critical transition of the rotating system from ref.~\cite{bradley_bose-einstein_2008}. Gray square: Mean field calculations. Even at low temperatures where we expect the Thomas--Fermi approximation used in~\cite{stringari_phase_1999} to be accurate, simulations predict a $\bar{\Omega}_c$ about 50\% of the analytical result.}
    \label{fig:vortex-novortex_diagram}
\end{figure}

In Fig.~\ref{fig:vortex-novortex_diagram}, we compare the transition located by CL simulation (filled orange circles) with the transition predicted by analytical analysis based on an approximate solution to the GPE extrapolated to finite temperature by Stringari~\cite{stringari_phase_1999} (solid blue line).
The result at $\mathrm{T}/\mathrm{T}_c^0 = 0$ from mean-field simulation is also plotted (filled gray square).

At $\mathrm{T} = 0$, the noise-free (mean-field) simulation (gray square) and analytical theory from ref.~\cite{stringari_phase_1999} (solid blue line) agree closely, likely because both approaches are solutions to the GPE in this limit.
At all temperatures investigated, the fluctuation-corrected transition obtained from CL simulations is located at about 50\% of the rotation frequency predicted by the approximate analytical theory.
Furthermore, the simulation data obtained do not seem to imply a smooth connection to the zero-temperature mean field result (gray square).

The apparent disconnect between the CL boundary and GPE result at $\mathrm{T}=0$ could be the result of a lack of data in the temperature range $0.0 < \mathrm{T} / \mathrm{T_c^0} < 0.1$, or could indicate that the shift of the transition to lower rotation frequency is induced by quantum fluctuations that would remain even as $\mathrm{T} \rightarrow 0$.
The relative independence of the shift as a function of temperature seems to suggest the later.

It seems unlikely that the disagreement between simulation and analytical theory above $\mathrm{T} = 0$ could be explained by a potential violation of the Thomas--Fermi assumption.
The kinetic energy of each simulation is reported in Tab.~\ref{tab:0-1transitions}.
Not only is the portion of internal energy due to kinetic energy highly dependent on simulation temperature despite the relatively constant shift in the critical transition, at the lowest temperature the kinetic energy is only 10\%--15\% of the total internal energy.
In previous studies of trapped gases, close agreement was achieved with Thomas--Fermi analytical results with similar kinetic energy fractions~\cite{simmons_thermodynamic_2023}.
A rigorous $\mathrm{T} = 0$ reference could further deconvolute the effects of thermal and quantum fluctuations.

\begin{table}
\caption{The location of $\bar{\Omega}_c$ predicted by our simulation method at various temperatures, given in both non-dimensional simulation units, and units of $\mathrm{T}_c^0$. We also calculate kinetic energy as a fraction of total internal energy for every simulation, and report the observed range across all $\bar{\Omega}$ for each temperature.} 
\label{tab:0-1transitions}
\begin{ruledtabular}
\begin{tabular}{cccc}
$\bar{\mathrm{T}}$ & $\mathrm{T}/\mathrm{T}_c^0$ & $\bar{\Omega}_c$ & $E_{\mathrm{KE}}$/$E$ \\ 
\hline
    Mean field & 0.43 & -- & --\\
    5	& 0.22 & 0.23 & 0.10--0.16\\
    10  & 0.45 & 0.22 & 0.20--0.26\\
    15  & 0.67 & 0.25 & 0.33--0.38\\
\end{tabular}
\end{ruledtabular}
\end{table}


Another possible source of error in the analytical theory is the use of ideal-gas results for the critical temperature and thermal depletion of the condensate.
CSCL simulation has been used to locate the critical temperature in an unstirred homogeneous gas ~\cite{delaney_numerical_2020}, and there exists an operator for the number of particles in the condensate phase which could be used to calculate depletion~\cite{G_K_book}.
We could therefore test the validity of Stringari's assumptions with CSCL simulation.
Investigation of the system around the critical temperature is highly compelling for this reason, and the fact that the vortex--no vortex and BEC--normal fluid boundaries approach each other, prompting questions about the nature of the transition.
Do the two boundaries continuously meet, as is suggested in Stringari's phase diagram, or is there a tricritical point at finite temperature and rotation speed?
However, the temperatures achieved here are too low to make any compelling statements about the validity of the assumptions regarding depletion and critical temperature or the nature of the transition.

However, further investigation of this higher temperature region is hindered by the difficulty of balancing simulation lifetime with vortex structure stability.
As temperature increases, the strength of fluctuations increases, requiring longer sampling times to obtain accurate operator estimates.
Due to the stochastic nature of the simulations, occasionally large excursions drive the system over energy barriers into other vortex configurations, or even destroy vortices entirely.
As simulation runtime is increased, so too is the likelihood of experiencing such a large excursion.
The result is that statistical robustness and stability become increasingly difficult to balance.
Because of the high cost of simulation in this region, investigation of the critical transition in the presence of vortices is left to future work. We note that the success of the CSCL method in locating the $\lambda$-transition of the unstirred homogeneous gas~\cite{delaney_numerical_2020} indicates there is no fundamental limitation of the method around the critical transition.

Finally, we discuss the exclusion of the purely imaginary noise in the $\bar \mu$ constraint equation.
As in the previous case, we calculate Helmholtz and internal energy at the minimum and maximum relevant rotation frequencies at the minimum and maximum temperatures studied both with and without the constraint noise.
In the present case, since the location of the transition is dependent on the difference in Helmholtz free energy between the no and one-vortex configurations, we run two sets of simulations at each rotation frequency: one seeded with a no-vortex solution and one seeded with a one-vortex solution.
The minimum and maximum rotation speeds are the lowest and highest values of $\bar \Omega$ used to interpolate the phase boundary at a given temperature.
For $\bar{\mathrm{T}} = 5$ this is $\bar \Omega = 0.200$ and $\bar \Omega = 0.250$, respectively, and for $\bar{\mathrm{T}} = 15$ it is $\bar \Omega = 0.225$ and $\bar \Omega = 0.300$.
As before, seeds are solutions of the GPE.
In all four cases, the difference in Helmholtz free energy between the full and approximate simulations was 0.1\% or less.

\section{Conclusion}
In this work, we have demonstrated the application of field-theoretic simulations based on the coherent-states, imaginary-time path integral representation of rotating  Bose-Einstein condensates.
We use complex Langevin sampling to handle the sign problem innate to the coherent-states representation and the rotation term in the Hamiltonian.
We leverage the convenience of field operators to generate high-resolution real-space images of BEC cloud density and momentum vector fields, as well as to calculate free energies of five different types of vortex structures as a function of rotation speed at finite temperature.
Finally, we map the phase diagram of the trapped, rotating BEC and find that the addition of thermal and quantum fluctuations at finite temperature stabilizes vortices and lowers the critical rotation frequency for vortex formation, $\bar{\Omega}_c$.
We speculate the effect is mainly due to quantum fluctuations, and provide evidence to support this hypothesis.

This work opens the way for further application of complex Langevin simulations to rotating systems of BECs.
The coherent states basis is flexible, and allows for the calculation of a wide variety of quantities which lend themselves to future studies.
For example, with field operators for current and vorticity, as well as the operators detailed in this work, one could elucidate further details about the BEC to normal fluid transition, particularly at rotation speeds near the initiation of vortex structures and regarding the possibility of a tricritical point between the BEC without a vortex, BEC with a single vortex, and normal fluid phases at finite rotation speed. A field-theoretic extension of the path integral ground state (PIGS) method\cite{Yan_2017} could also be used to distinguish the effect of quantum and thermal fluctuations.

More broadly, the ability to simulate bosonic many-body systems with near-linear scaling of computational cost with system volume and near independence with particle number, especially in models with a sign problem, promises new insights into problems for which current Monte Carlo-based simulation methods have struggled; for example artificial gauge fields applied to bosons on lattices~\cite{powell_bogoliubov_2011}, spinor gases~\cite{su_crystallized_2012, cabedo_excited-state_2021} and pseudo-spin-1/2 gases~\cite{stanescu_spin-orbit_2008}.

\begin{acknowledgments}
We thank Matthew Fisher, David Weld, and David Hall for helpful discussions.
K.K.\ acknowledges support from the National Science Foundation Graduate Research Fellowship under Grant No. 1650114.
G.H.F.\ and K.T.D.\ acknowledge
support from NSF DMR-2104255 for the theoretical method development.
Use was made of the BioPACIFIC Materials Innovation Platform computing resources of the National Science Foundation Award No. DMR-1933487, as well as computational facilities purchased with funds from the National Science Foundation (CNS-1725797) and administered by the Center for Scientific Computing (CSC). The CSC is supported by the California NanoSystems Institute and the Materials Research Science and Engineering Center (MRSEC; NSF DMR-2308708) at UC Santa Barbara.
\end{acknowledgments}

\bibliography{citations}

\end{document}